\documentclass[12pt]{article}
\usepackage{graphicx}

\def \bea{\begin{eqnarray}}
\def \beq{\begin{equation}}
\def \eea{\end{eqnarray}}
\def \eeq{\end{equation}}

\def \vert{\rule[-1mm]{0.2mm}{5mm}}

\begin{document}
\rightline{EFI 08-24}
\rightline{arXiv:0809.xxxx}
\rightline{September 2008}
\bigskip
\centerline{\bf M2 SIGNATURES IN $\psi(2S)$ RADIATIVE DECAYS}
\bigskip

\centerline{Jonathan L. Rosner\footnote{rosner@hep.uchicago.edu}}
\centerline{\it Enrico Fermi Institute and Department of Physics}
\centerline{\it University of Chicago, 5640 S. Ellis Avenue, Chicago, IL 60637}

\begin{quote}
The sensitivity of observables in radiative decays $\psi(2S) \to \gamma
\chi_{c1,2} \to \gamma \gamma J/\psi$ to quadrupole (M2) admixtures in the
dominant electric dipole (E1) transitions is explored.  Emphasis is placed on
distributions in a single angle, and several examples are given.
\end{quote}

\leftline{PACS numbers: 14.40.Gx, 13.20.Gd, 13.40.Hq, 12.39.Jh}
\bigskip

Radiative decays of charmonium ($c \bar c$) states are expected to be
dominated by electric dipole (E1) transitions, with higher multipoles
suppressed by powers of photon energy divided by quark mass \cite{E1dom}.
(An early discussion of charmonium radiative transitions may be found in
Ref.\ \cite{Jackson:1976mt}.)  The search for contributions of higher
multipoles is of interest as a source of information on the charmed quark's
magnetic moment, which is difficult to measure directly because of the
extremely short lifetime of charmed hadrons.  The possibility of anomalous
magnetic moments of heavy quarks being larger than those of light ones,
as a result of light-quark loops, was raised in Ref.\ \cite{Geffen:1980ex}.

Expectations for the magnitude of magnetic quadrupole (M2) transitions in
charmonium were calculated in Ref.\ \cite{Karl:1980wm}, and detailed
discussions of various ways to extract the M2/E1 ratio were presented in
Ref.\ \cite{Karl:1975jp}.  Many of these methods involved the use of
distributions in several decay angles.  The present note is devoted to
several sources of information on the M2/E1 ratio based on distributions
with respect to a single angle.  These distributions can be examined
in new data on $\psi(2S)$ decays from the CLEO Detector at the
Cornell Electron-Positron Storage Ring (CESR) \cite{CLEO}.

The distributions which will be examined are with respect to the following
angles:  (1) the angle $\theta'$ between the initial positron $e^+$ and
the photon $\gamma'$ in the $\psi(2S)$ center-of-mass system (c.m.s.), where
$\psi(2S) \to \gamma' \chi_c$; (2) the angle $\theta$ between the photon
$\gamma$ in $\chi_c \to \gamma J/\psi$ and the final $\ell^+$ direction in
$J/\psi \to \ell^+ \ell^-$, in the $J/\psi$ c.m.s.; (3) the angle
$\theta_{\gamma' \gamma}$ between $\gamma'$ and $\gamma$ in the cascade
$\psi(2S) \to \gamma' \chi_c \to \gamma' \gamma J/\psi$, evaluated in the
$\chi_c$ c.m.s., and (4) the angle $\theta_{\gamma' P}$ in the cascade
$\psi(2S) \to \gamma' \chi_c \to \gamma' P^+ P^-$, where $P = \pi$ or $K$,
between $\gamma'$ and $P^+$, evaluated in the $\chi_c$ c.m.s.  In cases (1)
through (3), one obtains information on the M2/E1 admixture from $\chi_{c1}$
and $\chi_{c2}$, while in case (4) only $\chi_{c2}$ provides information, as
$\chi_{c1}$ does not decay to $P^+ P^-$.  The transitions $\psi(2S) \to \gamma'
\chi_{c0}$ and $\chi_{c0} \to \gamma J/\psi$ are purely electric dipole (E1)
and do not provide information on the M2 amplitude.

It is conventional to define normalized amplitudes $a_i^J$ for the transitions
$\chi_{cJ} \to \gamma J/\psi$ and $b_i^J$ for the transitions $\psi(2S) \to
\gamma' \chi_{cJ}$ such that $\sum_i|a_i^J|^2 = \sum_i|b_i^J|^2 = 1$, with
$i=1,2,3$ corresponding to E1, M2, and E3
amplitudes, respectively.  We shall neglect possible E3 contributions,
which could only enter in the case of $\chi_{c2}$ intermediate states.
The current state of experimental and theoretical information on M2
admixtures is summarized in Table \ref{tab:m2}.  Energies for the
transitions $\psi(2S) \to \gamma' \chi_{cJ}$ and $\chi_{cJ} \to \gamma
J/\psi$ are summarized in Table \ref{tab:ens}.

\begin{table}
\renewcommand{\arraystretch}{1.3}
\caption{Magnetic quadrupole admixtures $b_2^J$ in $\psi(2S) \to \gamma'
\chi_{cJ}$ and $a_2^J$ in $\chi_{cJ} \to \gamma J/\psi$ from Crystal Ball
\cite{Oreglia:1981fx}, Fermilab E-760 \cite{Armstrong:1993fk}, Fermilab E-835
\cite{Ambrogiani:2001jw}, BES-II \cite{Ablikim:2004qn}, and theory with $m_c =
1.5$ GeV/$c^2$
\cite{Karl:1980wm}.
\label{tab:m2}}
\begin{center}
\begin{tabular}{c c c c c} \hline \hline 
Experiment & $b_2^{J=1}$ & $a_2^{J=1}$ & $b_2^{J=2}$ & $a_2^{J=2}$ \\ \hline
Crystal Ball & $0.077^{+0.050}_{-0.045}$ & $-0.002^{+0.008}_{-0.020}$
 & $0.132^{+0.098}_{-0.075}$ & $-0.333^{+0.116}_{-0.292}$ \\
E-760 & -- & -- & -- & $-0.14 \pm 0.06$ \\ 
E-835 & -- & 0.002$\pm$0.032$\pm$0.004 & --
 & $-0.093^{+0.039}_{-0.041}$$\pm$0.046 \\
BES-II & -- & -- & $-0.051^{+0.054~a}_{-0.036}$ & \\
Theory & $0.029(1+\kappa)$ & $-0.065(1+\kappa)$ & $0.029(1+\kappa)$
& $-0.096(1+\kappa)$ \\ \hline \hline
\end{tabular}
\end{center}
\leftline{$^a$ Result of fit with octupole moment $b_3 =
 -0.027^{+0.043}_{-0.029}$.}
\end{table}

\begin{table}
\renewcommand{\arraystretch}{1.3}
\caption{Energies of photons in $\psi(2S) \to \gamma' \chi_{cJ}$ and
$\chi_{cJ} \to \gamma J/\psi$, evaluated in rest frame of decaying particle,
in MeV \cite{PDG}.
\label{tab:ens}}
\begin{center}
\begin{tabular}{c c c} \hline \hline
$J$ & $E_{\gamma'}$ & $E_{\gamma}$ \\ \hline
 0  & $261.35 \pm 0.33$ & $303.05 \pm 0.32$ \\
 1  & $171.26 \pm 0.07$ & $389.36 \pm 0.07$ \\
 2  & $127.60 \pm 0.09$ & $429.63 \pm 0.08$ \\ \hline \hline
\end{tabular}
\end{center}
\end{table}

Helicity amplitudes $A_{|\nu|}^J$ for $\chi_{cJ} \to \gamma J/\psi$ decays
and $B_{|\nu|}^J$ for $\psi(2S) \to \gamma' \chi_{cJ}$ are related to
respective multipole amplitudes $a^J_{J_\gamma}$ and $b^J_{J'_\gamma}$
\cite{Karl:1975jp} as
\bea
A_{|\nu|}^J & = & \sum_{J_\gamma} a^J_{J_\gamma}
 \left( \frac{2J_\gamma+1}{2J+1} \right)^{1/2}
 (J_\gamma,1,1,|\nu|-1 ~\vert~ J,|\nu|)~, \\
B_{|\nu|}^J & = & \sum_{J'_\gamma} b^J_{J'_\gamma}
 \left( \frac{2J'_\gamma+1}{2J+1} \right)^{1/2}
 (J'_\gamma,1,1,|\nu'|-1 ~\vert~ J,|\nu'|)~.
\eea

Specifically, the transformations for the $A$ amplitudes are

\bea
J = 1:~~~~~~~~~A_1^{J=1} & = & \sqrt{\frac{1}{2}} a_1^{J=1}
                              -\sqrt{\frac{1}{2}} a_2^{J=1}~, \\
               A_0^{J=1} & = & \sqrt{\frac{1}{2}} a_1^{J=1}
                              +\sqrt{\frac{1}{2}} a_2^{J=1}~.
\eea
\bea
J = 2:~~~A_2^{J=2} & = & \sqrt{\frac{3}{5}}  a_1^{J=2}
                        -\sqrt{\frac{1}{3}}  a_2^{J=2}
                        +\sqrt{\frac{1}{15}} a_3^{J=2}~,\\
         A_1^{J=2} & = & \sqrt{\frac{3}{10}} a_1^{J=2}
                        +\sqrt{\frac{1}{6}}  a_2^{J=2}
                        -\sqrt{\frac{8}{15}} a_3^{J=2}~,\\
         A_0^{J=2} & = & \sqrt{\frac{1}{10}} a_1^{J=2}
                        +\sqrt{\frac{1}{2}}  a_2^{J=2}
                        +\sqrt{\frac{2}{5}}  a_3^{J=2}~,
\eea
with a similar set of transformations for the $B$ amplitudes.

Taking the charmed quark to have a mass $m_c = 1.5$ GeV/$c^2$ and an anomalous 
magnetic moment $\kappa$, noting the photon energies in Table \ref{tab:ens},
and approximating $b_1^J\simeq a_1^J\simeq 1$, one predicts \cite{Karl:1980wm}
\bea
b_2^{J=1} & = & \frac{E_{\gamma'}[\psi(2S) \to \gamma' \chi_{c1}]}{4 m_c}
 (1 + \kappa) = 0.029(1 + \kappa)~,\\
a_2^{J=1} & = & - \frac{E_{\gamma}[\chi_{c1} \to \gamma J/\psi]}{4 m_c}
 (1 + \kappa) = -0.065(1 + \kappa)~,
\eea
\bea
b_2^{J=2} & = & \frac{3}{\sqrt{5}} \frac{E_{\gamma'}[\psi(2S) \to \gamma'
 \chi_{c2}]}{4 m_c} (1 + \kappa) = 0.029(1 + \kappa)~,\\
a_2^{J=2} & = & - \frac{3}{\sqrt{5}} \frac{E_{\gamma}[\chi_{c2} \to \gamma
 J/\psi]}{4 m_c} (1 + \kappa) = -0.096(1 + \kappa)~.
\eea
These predictions are summarized in Table \ref{tab:m2}.

Although the most recent measurements, those of Fermilab E-835, are
consistent with predictions, no convincing signal for any M2 admixture has
yet been obtained.  Prediction of the {\it ratios} $a_2^{J=2}/a_1^{J=1}$
and $b_2^{J=2}/b_1^{J=1}$ are likely to be more reliable than individual
predictions because the charmed quark mass and anomalous moment cancel in
these ratios:
\beq
a_2^{J=2}/a_1^{J=1} = \frac{3}{\sqrt{5}}\frac{429.63}{389.36} = 1.48~,~~
b_2^{J=2}/b_1^{J=1} = \frac{3}{\sqrt{5}}\frac{127.60}{171.26} = 1.00~.
\eeq

We introduce a shorthand notation
\bea x_J & = &   E_{\gamma}[\chi_{cJ} \to J/\psi] (1 + \kappa)/(4 m_c)~,\\
  x'_J & = & - E_{\gamma'}[\psi(2S) \to \gamma' \chi_{cJ}](1 + \kappa)
 /(4 m_c)~,
\eea
in terms of which the helicity
amplitudes for $\chi_{cJ} \to \gamma J/\psi$ have the relative values (with
arbitrary normalization for overall rates) \cite{Karl:1980wm}
\beq
A_0^{J=0} = \sqrt{2}~,
\eeq
\beq
A_1^{J=1} = \sqrt{3}(1 + x_1)~,~~
A_0^{J=1} = \sqrt{3}(1 - x_1)~,
\eeq
\beq
A_2^{J=2} = \sqrt{6}(1 + x_2)~,~~
A_1^{J=2} = \sqrt{3}(1 - x_2)~,~~
A_0^{J=2} = 1 - 3x_2~,
\eeq
\beq
B_0^{J=0} = \sqrt{2}~,
\eeq
\beq
B_1^{J=1} = \sqrt{3}(1 + x'_1)~,~~
B_0^{J=1} = \sqrt{3}(1 - x'_1)~,
\eeq
\beq
B_2^{J=2} = \sqrt{6}(1 + x'_2)~,~~
B_1^{J=2} = \sqrt{3}(1 - x'_2)~,~~
B_0^{J=2} = 1 - 3x'_2~,
\eeq
Many observables are most simply expressed in terms of the $A$s and $B$s
\cite{Karl:1975jp}

We begin with the distributions with respect to the angles $\theta'$ and
$\theta$ defined above.  They are:
\bea
J=0: W(\cos \theta) & \propto & 1 + \cos^2 \theta~, \\
J=1: W(\cos \theta) & \propto & \frac{1 + \cos^2 \theta}{2} (A_0)^2 
 + \sin^2 \theta (A_1)^2~, \\
J=2: W(\cos \theta) & \propto & \frac{1 + \cos^2 \theta}{2} [(A_2)^2+(A_0)^2]
 + \sin^2 \theta (A_1)^2~.
\eea
with similar expressions for $\theta \to \theta',~A_\nu \to B_\nu$.  We use
the expressions for helicity amplitudes, keep terms to first order in $x_J$,
define $z = \cos \theta$, and normalize $\int_{-1}^1 W(z) dz = 1$.  The
results are
\bea
J=1: W(z) & = & \frac{3}{16}\left[3-z^2+2x(1-3z^2)\right]~, \label{eqn:wz1}\\
J=2: W(z) & = & \frac{3}{80}\left[13+z^2-6x(1-3z^2)\right]~ \label{eqn:wz2},
\eea
and similarly for primed quantities.  The terms linear in $x$ give zero when
integrated over the full range of $z$, as expected because the M2--E1
interference term should not contribute to the decay rate.

\begin{figure}
\mbox{\includegraphics[width=0.48\textwidth]{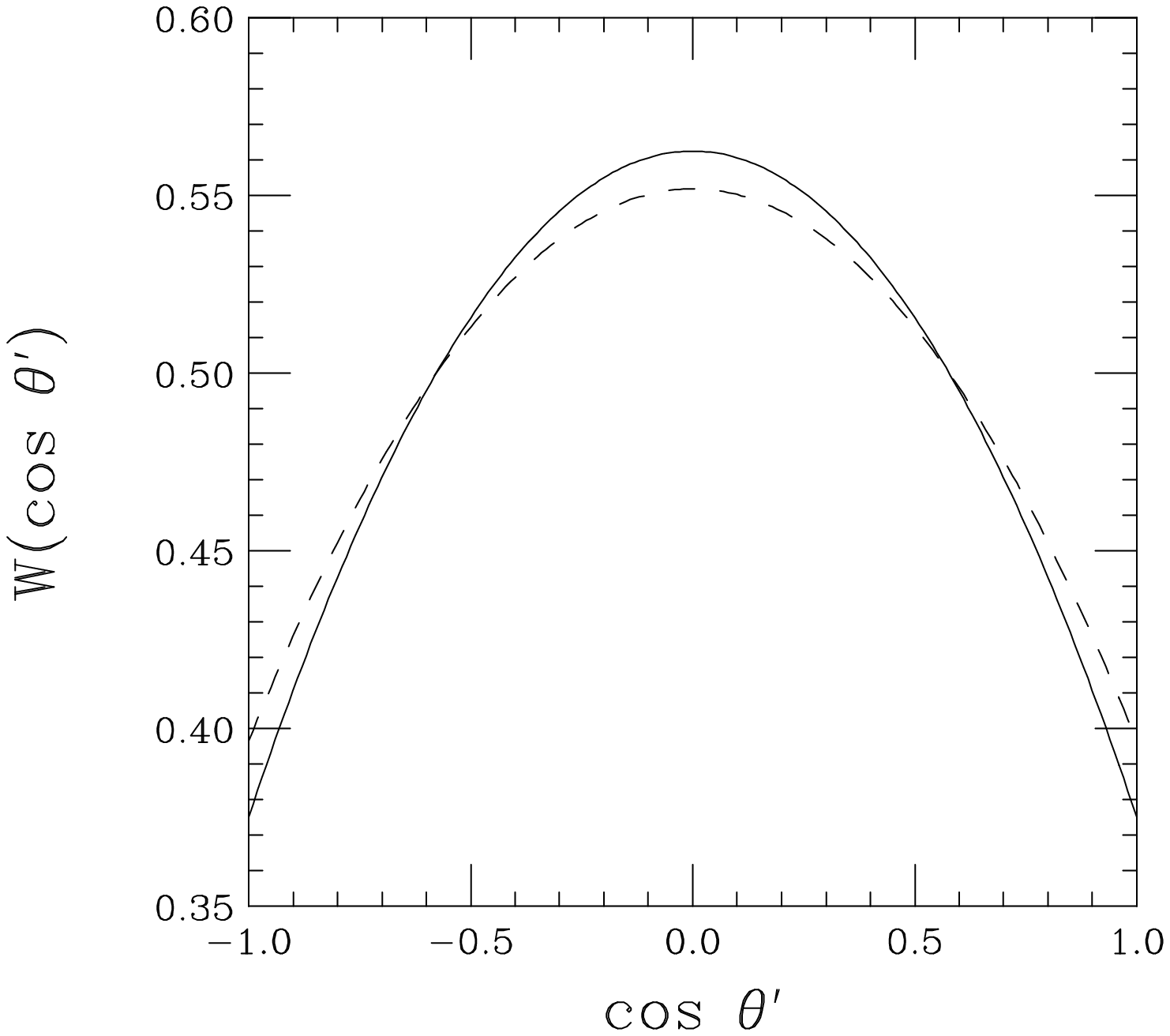}
      \includegraphics[width=0.48\textwidth]{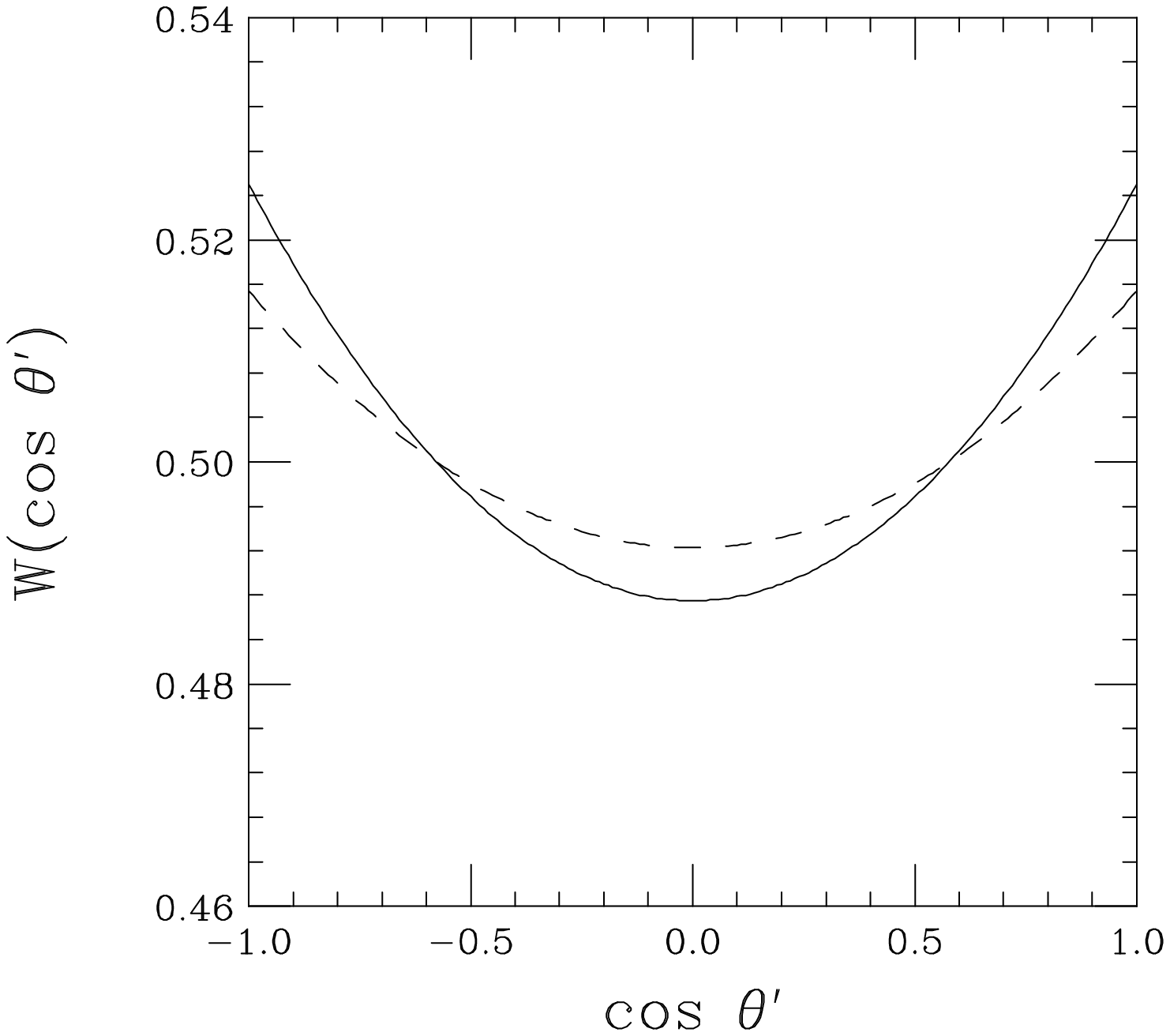}}
\caption{Distributions in $\cos \theta'$, angle between photon $\gamma'$
and beam axis in $e^+ e^- \to \psi(2S) \to \gamma' \chi_{cJ}$.  Solid lines:
pure E1; dashed lines: including expected M2 admixture.  Left:  $J=1$, $x'_1
= -0.029$; right: $J=2$, $x'_2 = -0.021$.
\label{fig:m2cp}}
\end{figure}

For the transitions $\psi(2S) \to \gamma' \chi_{c1,2}$, one expects
\beq
x'_1 = - b_1^{J=1} = -0.029~,~~
x'_2 = - \frac{\sqrt{5}}{3}b_1^{J=1} = -0.021~.
\eeq
The expected angular distributions for the above $x'_J$ values are compared
with those for $x'_J = 0$ in Fig.\ \ref{fig:m2cp}.  The expected M2 admixtures
lead to distributions which are very slightly flatter in $\theta'$ than the
pure E1 transitions.

\begin{figure}
\mbox{\includegraphics[width=0.48\textwidth]{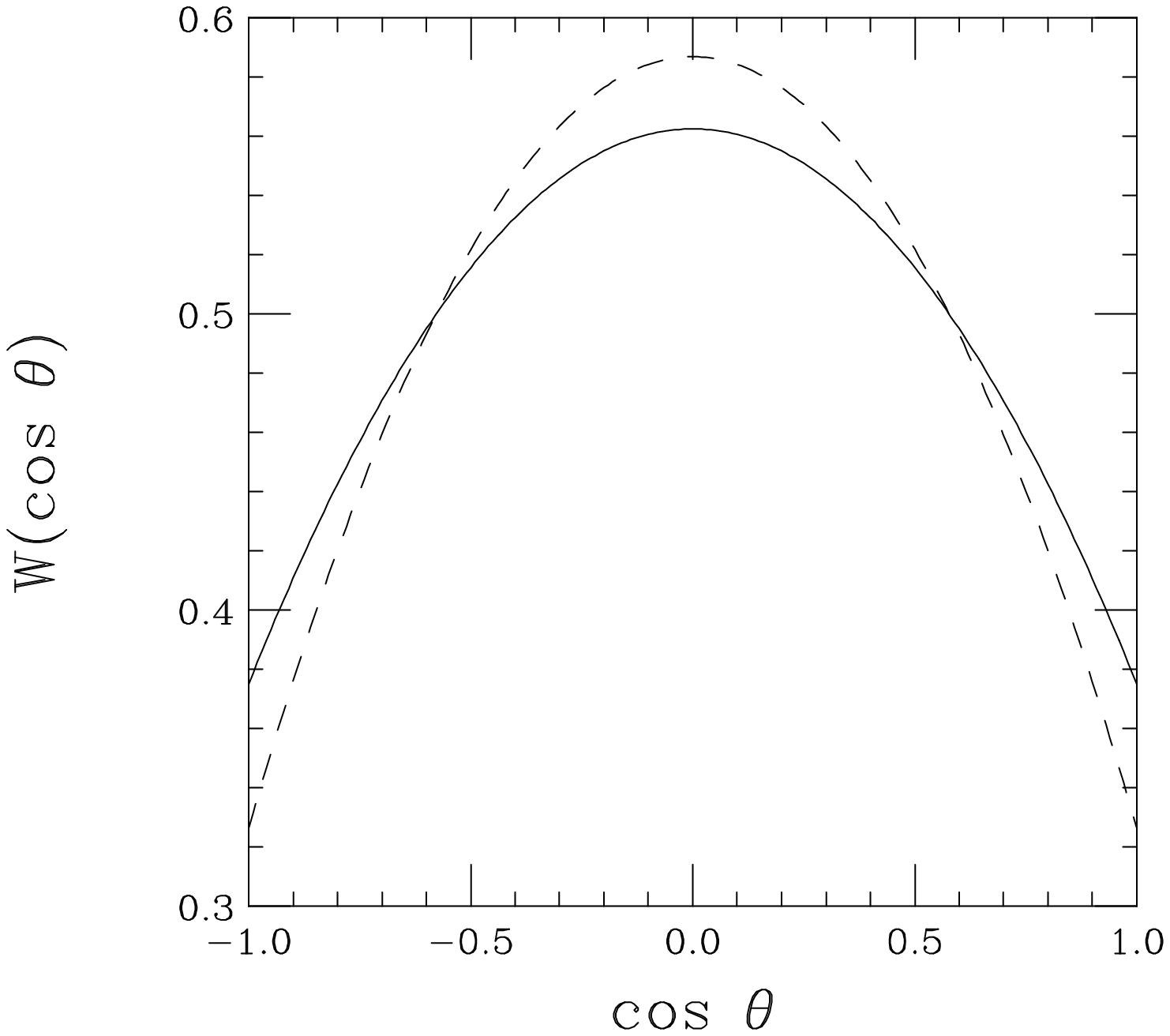}
      \includegraphics[width=0.48\textwidth]{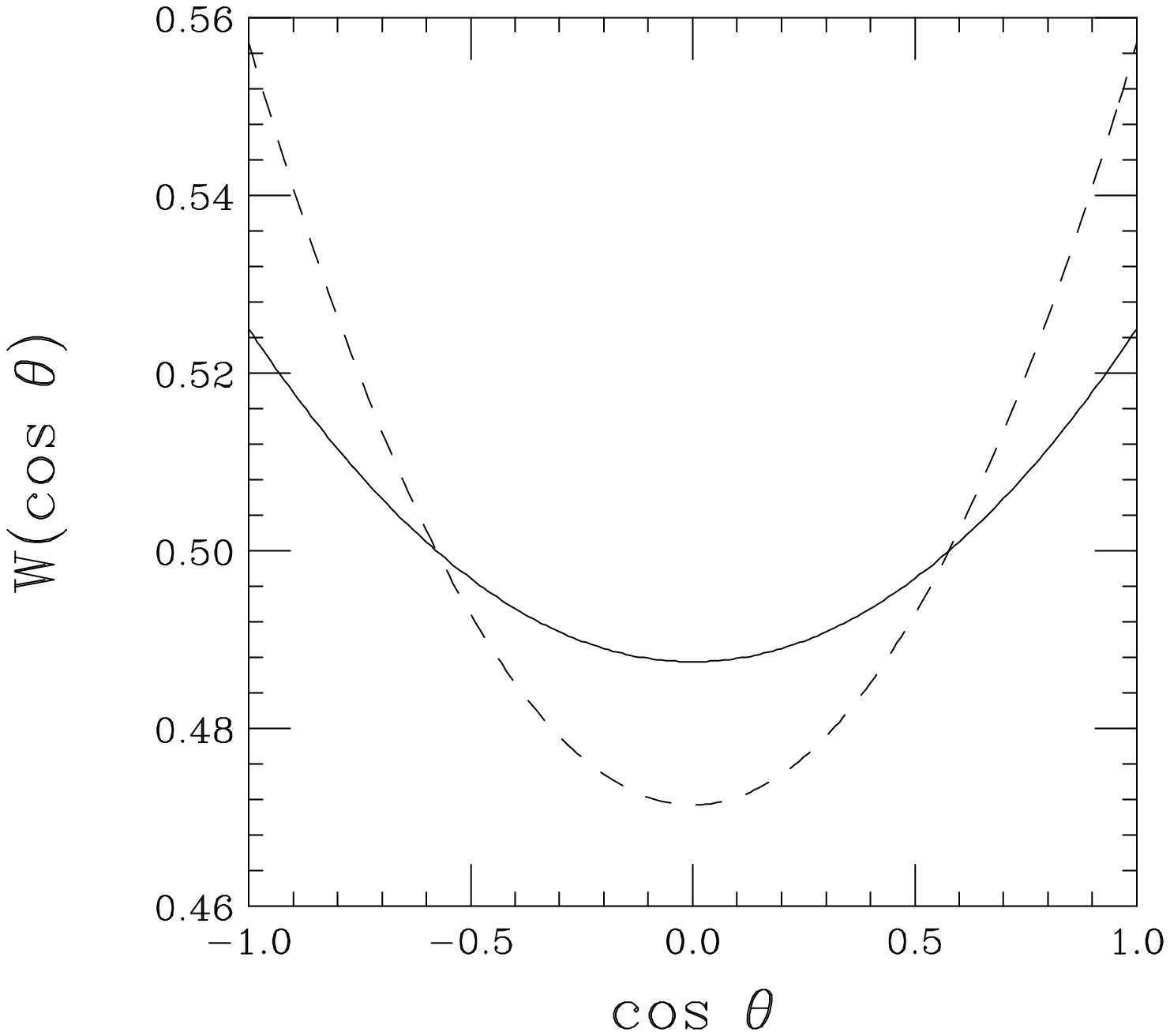}}
\caption{Distributions in $\cos \theta$, angle between photon $\gamma$ and
lepton pair axis in $\chi_{cJ} \to \gamma J/\psi \to \gamma e^+ e^-$, evaluated
in $J/\psi$ rest frame.  Solid lines: pure E1; dashed lines: including expected
M2 admixture.  Left:  $J=1$, $x_1 = 0.065$; right: $J=2$, $x_2 = 0.072$.
\label{fig:m2c}}
\end{figure}

For the transitions $\chi_{c1,2} \to \gamma J/\psi$ the photon energies are
larger and the signs of the M2 contributions are such that the distributions
in $\theta$ are slightly more pronounced than for pure E1 transitions.  These
distributions are shown in Fig.\ \ref{fig:m2c}.

\begin{figure}
\mbox{\includegraphics[width=0.48\textwidth]{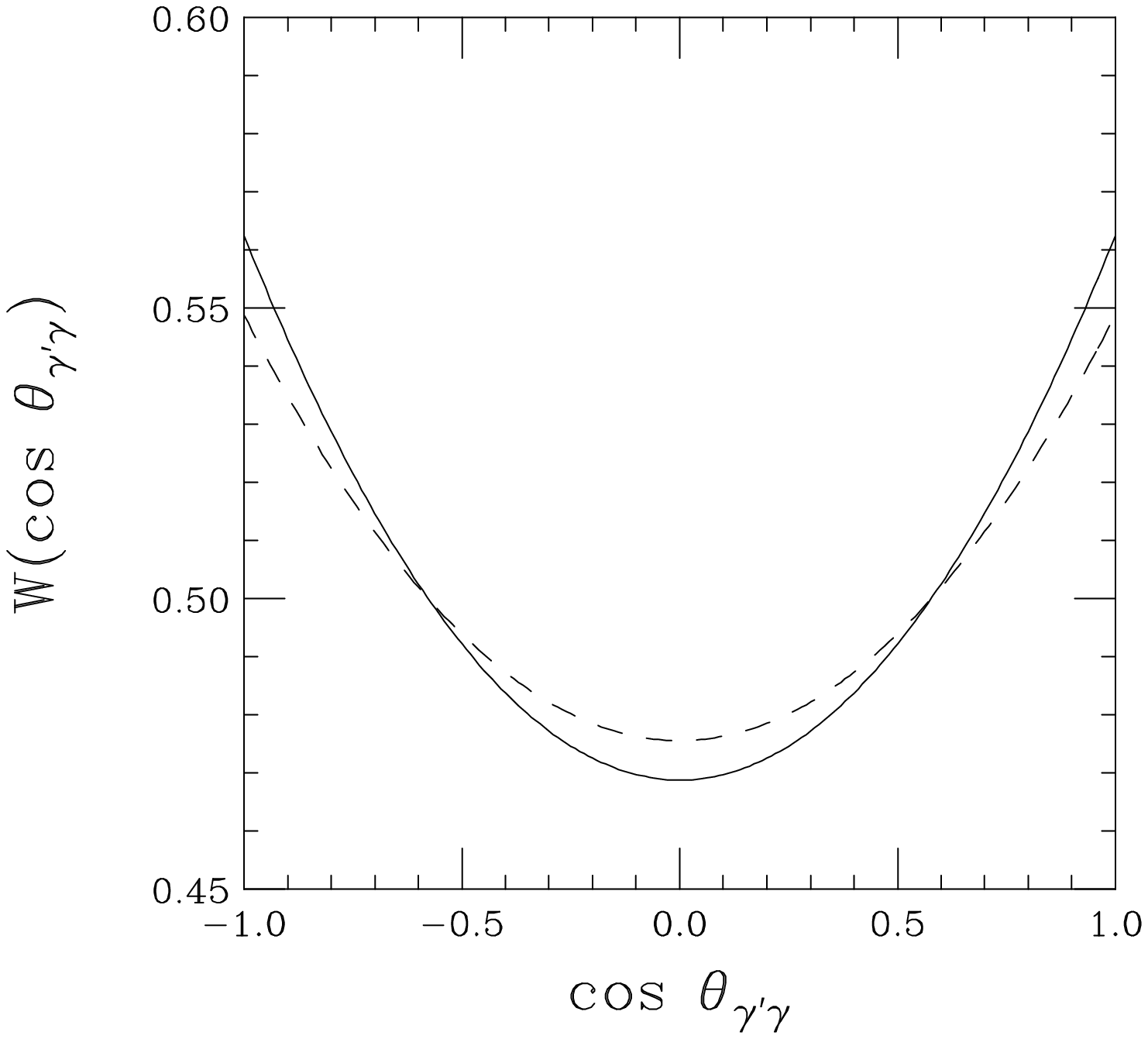}
      \includegraphics[width=0.48\textwidth]{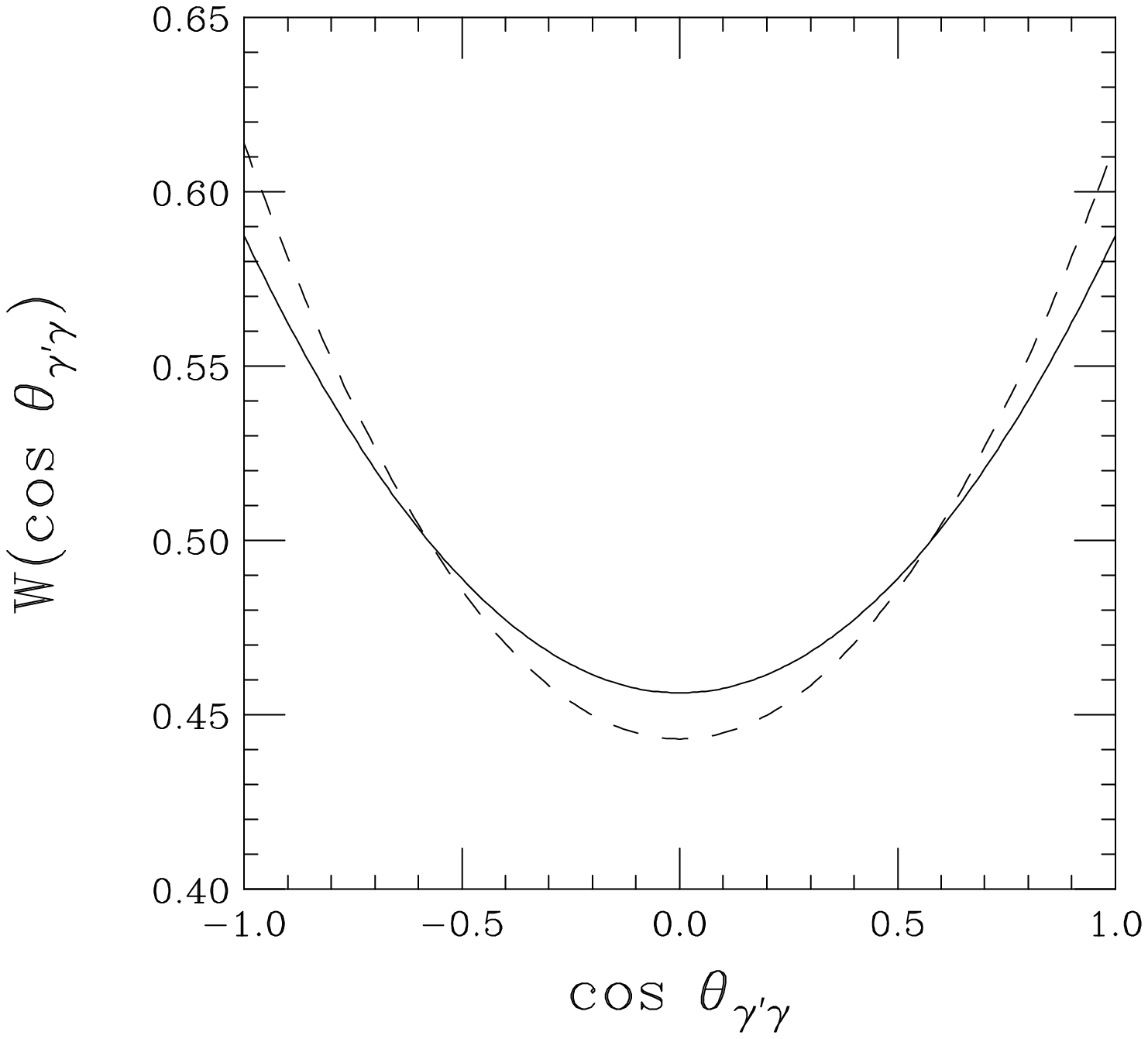}}
\caption{Distributions in $\cos \theta_{\gamma' \gamma}$, angle between photon
$\gamma$ in $\chi_{cJ} \to \gamma J/\psi$ and $\gamma'$ in $\psi(2S) \to
\gamma' \chi_{cJ}$, evaluated in $\chi_{cJ}$ rest frame.  Solid lines: pure E1;
dashed lines: including expected M2 admixture.  Left:  $J=1$, $x_1 + x'_1 =
0.036$; right: $J=2$, $x_2 + x'_2 = 0.050$.
\label{fig:gg}}
\end{figure}

We next discuss the effect of an M2 admixture on the photon-photon correlation
angle $\theta_{\gamma' \gamma}$.  Defining $z_c \equiv \cos \theta_{\gamma'
\gamma}$, the distributions in terms of helicity amplitudes are
\cite{Karl:1975jp}:
\bea
J=1: W(z_c) & \propto & (A_1 B_1)^2 + 2 (A_1 B_0)^2 + 2 (A_0 B_1)^2 + z_c^2
 (2 A_0^2 - A_1^2)(2 B_0^2 - B_0^2)~,\\
J=2: W(z_c) & \propto & (A_2 B_2)^2 + 4(A_2 B_1)^2 + 4 (A_1 B_2)^2
 + 6 (A_2 B_0)^2 + 6 (A_0 B_2)^2 + 4 (A_1 B_1)^2 
 \nonumber \\
 & + & 4(A_0 B_0)^2 + 6 z_c^2[(A_2^2 - 2 A_0^2)(B_2^2 - 2 B_0^2)
 - 2 (A_1^2 - 2 A_0^2)(B_1^2 - 2 B_0^2)] \nonumber \\
 & + & z_c^4 (A_2^2 - 4 A_1^2 + 6 A_0^2)(B_2^2 - 4 B_1^2 + 6 B_0^2)
\eea
Keeping terms to first order in $x_J$ and $x'_J$, one finds that the
$z_c^4$ terms cancel and
\bea
J=1: W(z_c) & = & \frac{3}{32}[5 + z_c^2 + 2(x_1 + x'_1)(1 - 3z_c^2)]~, \\
J=2: W(z_c) & = & \frac{1}{160}[73 + 21 z_c^2 + 42(x_1 + x'_1)(3z_c^2-1)]~.
\eea
Here we have normalized $\int_{-1}^1 dz_c W(z_c) = 1$.  The angular
distributions for $J=1$ and $J=2$ are shown in Fig.\ \ref{fig:gg}.

As for the single-angle distributions, the deviation from pure E1 behavior is
proportional to a Legendre polynomial $P_2(z_c)$.  The deviations are of
opposite signs for $J=1$ and $J=2$, and slightly larger for $J=2$ as a result
of the larger value of $x_J + x'_J$.

\begin{figure}
\begin{center}
\includegraphics[width=0.75\textwidth]{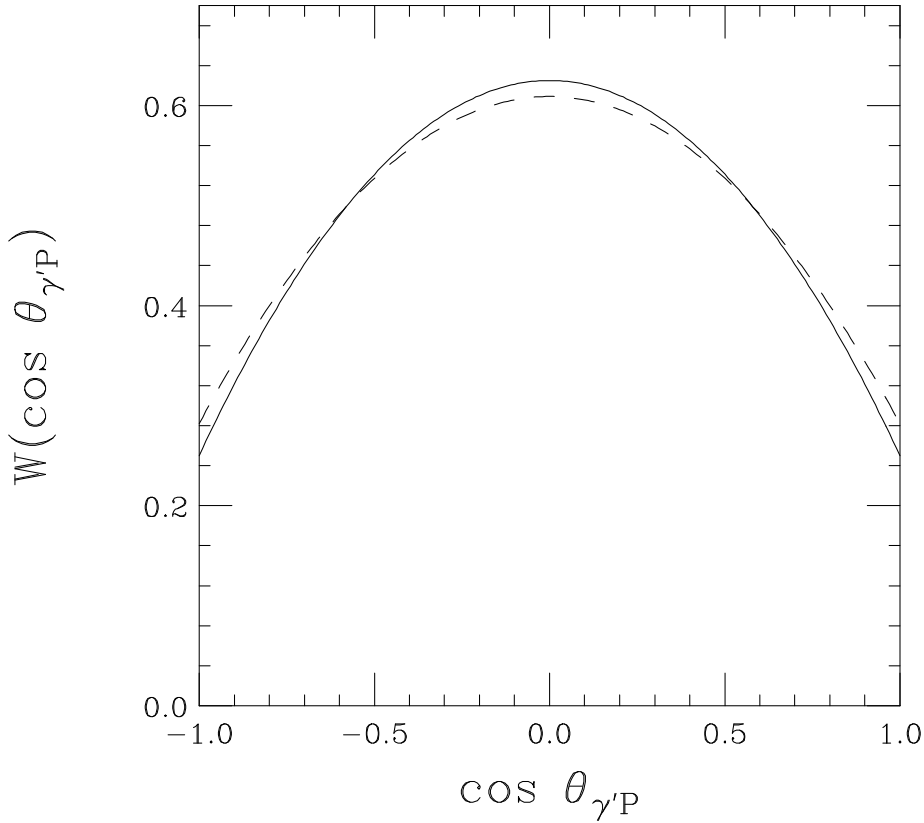}
\end{center}
\caption{Distribution in $\cos_{\theta' P}$, where $\theta_{\gamma' P}$ is the
angle between the photon $\gamma'$ in $\psi(2S) \to \gamma \chi_{c2}$ and the
pseudoscalar meson $P^+$ in the decay $\chi_{c2} \to P^+ P^-$.  Solid curve:
pure E1; dashed curve:  with expected M2 admixture corresponding to $x'_2
= -0.021$.
\label{fig:gp}}
\end{figure}

We conclude with a discussion of the effect of an M2 admixture on the angle
$\theta_{\gamma' P}$ between the photon $\gamma'$ in $\psi(2S) \to \gamma
\chi_{c2}$ and the pseudoscalar meson $P^+$ in the decay $\chi_{c2} \to
P^+ P^-$, where, for example, $P = \pi$ or $K$.  This distribution is
given in terms of $z_1 \equiv \cos \theta_{\gamma' P}$ and the helicity
amplitudes $B$ by \cite{Karl:1975jp}
\beq
W(z_1) \propto 3(B_2)^2 (1 - z_1^2)^2 + 12(B_1)^2(1 - z_1^2)z_1^2
 + 2 (B_0)^2 (3 z_1^2 - 1)^2~.
\eeq
Substituting for the helicity amplitudes, keeping terms of order $x'_2$,
and normalizing \\ $\int_{-1}^1 dz_1 W(z_1) = 1$, one finds
\beq
W(z_1) = \frac{5 - 3 z_1^2}{8} + \frac{3}{4} x'_2 (1 - 3 z_1^2)~.
\eeq
This expression is plotted for a pure E1 transition and for the expected
M2 admixture in Fig.\ \ref{fig:gp}.
An analysis by the BES-II Collaboration \cite{Ablikim:2004qn} based
on 14 million $\psi(2S)$ decays finds no evidence for higher multipoles
(see Table \ref{tab:m2}).

The effects we have displayed all are rather small.  Based on the previous
experimental signatures summarized in Table \ref{tab:m2}, the most promising
distributions in which to see significant effects are probably Eqs.\
(\ref{eqn:wz1}) and (\ref{eqn:wz2}) as applied to the transition $\chi_{c1,2}
\to \gamma J/\psi$ (see Fig.\ \ref{fig:m2c}).  A simulation will be necessary
to tell whether the present CLEO data sample of some 27 million $\psi(2S)$
decays \cite{CLEO} is adequate to see such effects.

I thank K. Seth for the question which led to this investigation.  This work
was supported in part by
the United States Department of Energy through Grant No.\ DE-FG02-90ER-40560.

\end{document}